\documentclass[12pt]{article}
\usepackage{graphics}
\begin{document}
%------------------------------------------------------------------------------
\markright{Holonomy in Schwarzschild...}
%------------------------------------------------------------------------------
%-----------------------------------------------------------------------------
\title{Holonomy in the Schwarzschild-Droste Geometry}
\author{Tony Rothman~$^*$, George F.\ R.\ Ellis$^\dagger$, and Jeff
Murugan~$^\ddagger$
\\[2mm]
{\small\it \thanks{trothman@titan.iwu.edu}}~Dept.\ of Physics,
Illinois Wesleyan University, Bloomington, IL 61702, USA.\\
{\small \it\thanks{ellis@maths.uct.ac.za}}
{\small\it\thanks{jeff@hbar.mth.uct.ac.za}} Dept.\ of Maths and Applied
Maths,\\ University of Cape Town, Rondebosch 7700, Cape, South
Africa.\\ }
\date{{\small \LaTeX-ed \today}}
%-----------------------------------------------------------------------------
\maketitle
%----------------------------------------------------------------------

\begin{abstract}
\noindent
Parallel transport of vectors in curved spacetimes generally
results in a deficit angle between the directions of the initial
and final vectors. We examine such holonomy in the
Schwarzschild-Droste geometry and find a number of interesting
features that are not widely known.  For example, parallel
transport around circular orbits results in a quantized band
structure of holonomy invariance.  We also examine radial
holonomy and extend the analysis to spinors and to the
Reissner-Nordstr\"om metric, where we find qualitatively
different behavior for the extremal ($Q = M$) case.  Our
calculations provide a toolbox that will hopefully be useful in
the investigation of quantum parallel transport
in Hilbert-fibered spacetimes.

\vspace*{5mm} \noindent PACS: 04.20-q, 04.70Bw, 04.20-Cv\\
Keywords: Holonomy, Black holes, Schwarzschild.
\end{abstract}

%-----------------------------------------------------------------------------
\section{Introduction}
\label{sec1}
%-----------------------------------------------------------------------------

Holonomy transformations measure the change in direction acquired
by a vector under parallel transport around a closed loop, or
between two distinct points via different paths. To be more
precise, if $T_{p}$ is the tangent space to a manifold at the
point $p$, then a holonomy transformation is a set of linear maps
from $T_{p}$ into itself, induced by parallel transfer around
closed paths based on $p$. Each such path defines an element of
the holonomy group, determining the {\it deficit angle} between
the initial and final positions of a vector after such parallel
transport. It is a global property of the manifold and as such can
also serve as a tool for the {\it global} classification of
spacetimes in a manner similar to but distinct from the {\it
local} Petrov and Segre type classifications. In this regard, the
holonomy group structure of various (simply connected) spacetimes
has received exhaustive study {\cite{Hall2000}}. From the point of
view of its intrinsic mathematical interest as well, holonomy in
various spacetimes has been extensively examined, although in more
``exotic" settings, such as cylindrically symmetric spacetimes or
cosmic string backgrounds (see \cite{Bezerra96, Bezerra87} and
references therein).  Holonomy properties of one of the most
fundamental and important spacetimes, the
Schwarzschild-Droste\footnote {Johannes Droste, a pupil of
Lorentz, independently announced the ``Schwarzschild" exterior
solution within four months of Schwarzschild. See his ``The field
of a single centre in Einstein's theory of gravitation, and the
motion of a particle in that field," Koninklijke Nederlandsche
Akademie van Wetenschappen, Proceedings {\bf 19}, 197 (1917).}
geometry, have received virtually no attention in the literature.
To the best of our knowledge, only one paper \cite{Bollini}
reports any results in the Schwarzschild-Droste background, and
this as a special case of the rotating (Kerr) solution.\\

Apparently, the neglect of the Schwarzschild-Droste spacetime is
for a  simple reason: When asked whether a vector parallely
transported in an `equatorial' orbit around a Schwarzschild-Droste
(SD) black hole will manifest a deficit angle after completing a
full circle, most relativists answer, ``No."  The intuitive
response evidently relies on one's sense of spherical
symmetry---nothing changes during completion of the orbit---and on
the fact that around the equator of an ordinary two-sphere the
phase change is indeed zero. Here, however, is a striking case
where intuition fails. It is easy to show, as we do below, that
parallel transport in a circular orbit around a SD black hole
definitely results in a nonzero deficit angle---the vector has
changed direction. The first and most important point to be made
regarding holonomy in the SD geometry, therefore, is that nonzero
holonomy exists and consequently provides a gravitational analog
to the Aharonov-Bohm\footnote{ The Aharonov-Bohm effect was
actually predicted earlier by W.\ Ehrenberg and R.\ E.\ Siday,
Proc. Phys. Soc. London {\bf B62}, 8 (1949).} effect
\cite{AB59,Stachel1982}.  (Strictly speaking, the Aharonov- Bohm
effect takes place in a region where there is no electromagnetic
field, whereas the SD spacetime certainly contains a gravitational
field.  A closer analogy would be an asymptotically flat,
cylindrically symmetric space time, such as that produced by
cosmic strings \cite{Bezerra87}, which contains a conical
singularity.)\\

Beginning with the simple calculation for circular orbits, we explore
holonomy along a variety of paths in the SD geometry, for both geodesic and
non geodesic motion.
We find several surprising results.  Although a few of these are presented
in \cite{Bollini}, the majority are not and, in any case, none of them appear
to be widely known. We then carry out the calculations in loop variables,
extend the results to spinors and extremal Reissner-Nordstr\"om geometry,
and finally discuss some features that may bear on future calculations
involving ``quantum holonomy."

%-----------------------------------------------------------------------------
\section{Two-sphere and Schwarzschild-Droste metric}
\setcounter{equation}{0} \label{sec2}
%-----------------------------------------------------------------------------

It is an elementary exercise to show (see \cite{PB})
that parallel transporting a vector
 ${\bf A } = A^\phi {\bf {\hat e_\phi}} + A^\theta {\bf {\hat e_\theta}}$
along a constant $\theta$ curve on the surface of an ordinary two-sphere,
leads to

\begin{eqnarray}
A^\phi & = &\frac{\alpha \cos(\phi\ \cos\theta)
            - \beta \sin(\phi\ \cos\theta)}{\sin\theta}\nonumber\\
A^\theta & = & \alpha \sin(\phi\ \cos\theta) + \beta \cos(\phi\ \cos\theta),
\label{globe0}
\end{eqnarray}

\noindent
where $\alpha$ and $\beta$ are integration constants.  Note that when $\phi = 0$
we have $A^\phi = \alpha/sin\theta$ and $A^\theta = \beta$, and that after
completion of a full circle

\begin{eqnarray}
A^\phi & = &\frac{\alpha \cos(2\pi\ \cos\theta)
            - \beta\sin(2\pi\ \cos\theta)}{\sin\theta}\nonumber\\
A^\theta & = & \alpha \sin(2\pi\ \cos\theta) + \beta \cos(2\pi\ \cos\theta).
\label{globe2pi}
\end{eqnarray}

\noindent We see that on the equator ($\theta = \pi/2$) or at the
north pole ($\theta = 0$), parallel transport has no effect on the
components of ${\bf A}$, but that for arbitrary $\theta$,
$A^\phi(2\pi)$ and $A^\phi(0)$ differ, similarly for
$A^\theta(2\pi)$ and $A^\theta(0)$. As mentioned, it is perhaps
the zero result on the equator that helps to lead one's intuition
astray and to predict that the same will hold for the SD geometry.
However, this is not the case. We begin by setting the scene in
which we will work. Let $({\cal M},{\bf g})$ be a four-dimensional
Lorentzian manifold with Reissner-Nordstr\"om
 metric. The line element is, as usual,

\begin{equation}
ds^2=-(1-\frac{2M}r + \frac{Q^2}{r^2})dt^2+(1-\frac{2M}r
        + \frac{Q^2}{r^2})^{-1}dr^2
            +r^2(d\theta ^2+\sin^2\theta d\phi ^2),
\end{equation}

\noindent
where $Q$ is the charge and $M$ the mass. Most of the paper will be concerned
solely with the SD case, which is obtained by setting $Q=0$.
In section \ref{RN}, however, we will want to compare a few features of
the Schwarzschild-Droste geometry with the Reissner-Nordstr\"om case so it
will prove useful to have the relevant quantities available. For convenience
we will carry out calculations in an orthonormal
tetrad. The obvious choice is defined by the dual 1-form basis
$\omega^a=\omega _{\;i}^a(x^j)dx^i$, where

\begin{eqnarray}
 \omega ^0 &=&\omega ^t=(1-\frac{2M}r + \frac{Q^2}{r^2})^{1/2}dt, \nonumber\\
 \omega ^1 &=&\omega ^r=(1-\frac{2M}r + \frac{Q^2}{r^2})^{-1/2}dr, \nonumber\\
 \omega ^2 &=&\omega ^\theta =r\,d\theta , \nonumber\\
 \omega ^3 &=&\omega ^\phi =r\sin \theta \,d\phi .
 \label{forms}
\end{eqnarray}

\noindent The connection forms are defined by $d\omega
_{\;}^a=-\omega _{\;b}^a(x^j)\wedge \omega _{\;}^b.$ For the above
1-forms, we can choose the connection forms \cite{Wald} as

\begin{eqnarray}
 \omega _{\;r}^t &=&(\frac M{r^2} - \frac{Q^2}{r^3}) (1-\frac{2M}r
             + \frac{Q^2}{r^2})^{-1/2}\omega ^t=\omega
 _{\;t}^r, \nonumber\\
 \omega _{\;r}^\theta &=&\frac 1r(1-\frac{2M}r
             + \frac{Q^2}{r^2})^{1/2}\omega ^\theta =-\omega
 _{\;\theta }^r\;, \nonumber\\
 \omega _{\;r}^\phi &=&\frac 1r(1-\frac{2M}r
             + \frac{Q^2}{r^2})^{1/2}\omega ^\phi =-\omega
 _{\;\phi }^r\;, \nonumber\\
 \omega _{\;\theta }^\phi &=&\frac{\cot \theta }r\omega ^\phi =-\omega
 _{\;\phi }^\theta .
 \label{connectionForms}
\end{eqnarray}

\noindent
Let ${\bf A} = A^{\mu}\partial_{\mu}$ be some vector field on ${\cal M}$ i.e.,
a section of the tangent bundle of ${\cal M}$. By requiring that $\bf A$ be a
parallel section of the tangent bundle we can write the parallel transport
equation in a coordinate-free way as
\begin{equation}
 dA^{\mu}+ \omega^{\mu}_{\beta}A^{\beta}=0.
 \label{ParallelTransport}
\end{equation}

\noindent
Using the results of eqs.(\ref{forms}) and (\ref{connectionForms}) we write
eq.(\ref{ParallelTransport}) (for the Reissner-Nordstr\"om geometry) in
components as

\begin{eqnarray}
 dA^{t}+(M/r^2 - Q^2/r^3)A^{r}dt=&0&,\nonumber\\
 dA^{r}+(M/r^2 - Q^2/r^3)A^{t}dt
 - (1-2M/r + Q^2/r^2)^{\frac{1}{2}}A^{\theta}d\theta\nonumber\\
 - (1-2M/r + Q^2/r^2)^{\frac{1}{2}}A^{\phi}d\phi =&0&,\nonumber\\
 dA^{\theta}+(1-2M/r + Q^2/r^2)^{\frac{1}{2}}A^{r}d\theta=&0&,\nonumber\\
 dA^{\phi}+(1-2M/r + Q^2/r^2)^{\frac{1}{2}}A^{r}d\phi =&0&.
 \label{TransComp}
\end{eqnarray}

\noindent
In what follows we consider some special curves along which ${\bf A}$ is
parallel transported. Unless explicitly stated to the contrary we will
restrict ourselves to the Schwarzschild solution for which $Q=0$.

%-----------------------------------------------------------------------------
\section{Circular orbits}
\setcounter{equation}{0} \label{Circular}
%-----------------------------------------------------------------------------

Due to the spherical symmetry of the SD solution, we may take any
circular orbit to be equatorial, i.e., $\theta = \pi/2$.  Then,
since $r = const$ for these orbits we have for the tangent vector
$X^\nu=(X^t,0,0,X^\phi ) = (dt/d\lambda,0,0,d\phi/d\lambda)$.
Parameterizing such curves by $\phi$ and assuming constant speed
with $\mu \equiv dt/d\phi > c^{-1}$ we find from
eqs.(\ref{TransComp}) with $d\theta/d\phi = 0$,

\begin{eqnarray}
 dA_{\;\ }^t/d\phi &+&(\frac M{r^2}A^r)\mu=0, \label{DAt}\\
 dA_{\;\ }^r/d\phi &+&\frac M{r^2}A^t\mu-(1-\frac{2M}r)^{1/2}A^\phi =0,
      \label{DAr}\\
 dA_{\;\ }^\theta /d\phi &=&0, \label{DAth}\\
 dA_{\;\ }^\phi /d\phi &+&(1-\frac{2M}r)^{1/2}A^r=0. \label{DAphi}
\end{eqnarray}

\noindent
Equation (\ref{DAth}) is of course trivial and merely reflects
the constancy of $A^\theta$. The other equations may be easily
integrated to give

\begin{eqnarray}
 A^r(\phi) &=&\alpha \sin \omega \phi +\beta \cos \omega \phi  \label{Ar}
     \\
 A^t(\phi ) &=&\frac{\mu M}{\omega r^2}(\alpha \cos \omega \phi -\beta \sin
 \omega \phi )+\gamma , \label{At} \\
 A^\phi (\phi ) &=&\frac 1\omega (1-\frac{2M}r)^{1/2}(\alpha \cos \omega \phi
 -\beta \sin \omega \phi )+\delta \label{Aphi},
 \end{eqnarray}

\noindent
where the ``frequency" $\omega$ is given by

\begin{equation}
 \omega ^2\equiv 1-\frac{2M}r-\mu^2\frac{M^2}{r^4},
 \label{varpi}
\end{equation}

\noindent
and$\gamma$ and $\delta$ are integration constants.
Note that at $r = 2M$, $\omega^2$ is negative, whereas for fixed $\mu^2M^2$,
$\omega^2 \rightarrow 1$ as $r \rightarrow \infty$.  Thus $\omega^2$ will
change sign at some $ r = r_{crit}$, which depends
on the dimensionless parameter $\mu^2M^2/r^4$.  The oscillatory solutions above
are valid in the region $r>r_{crit}$.
The constants $\gamma$ and $\delta$ are
not independent.  Substituting the solutions back into the differential
equations shows that $\delta = (1-2M/r)^{-1/2}(M\mu/r^2) \gamma$. Also
note that for $r < r_{crit}$, the
solutions are exponential, in which case
the trigonometric functions in (\ref{Ar}) are replaced by the corresponding
hyperbolic function. Eqs.
(\ref{At}) and (\ref{Aphi}), obtained by integrating (\ref{Ar}), must be changed
accordingly.

\subsection{Constant-time circles}

The surprising properties of holonomy in the SD geometry can most easily be
seen by examining constant-time orbits. In this case, $\mu = 0$ and so
$r_{crit} = 2M$. Thus $\omega^2$ is always positive and the oscillatory
solutions are relevant. Suppose $\phi = 0$ at the start.  After $n$ loops
$\phi = 2\pi n$. $A^t = \gamma$ always and $A^\theta = constant$.
The remaining two components of $\bf A$ are

\begin{eqnarray}
 A_{\ }^\phi (2n\pi )&=&\alpha \cos [(1-\frac{2M}r)^{1/2}2n\pi ]-\beta \sin
 [(1-\frac{2M}r)^{1/2}2n\pi ],\label{Arconst}\\
 A_{\;\ }^r(2n\pi )&=&\alpha \sin [(1-\frac{2M}r)^{1/2}2n\pi ]+\beta \cos[(1-%
 \frac{2M}r)^{1/2}2n\pi ].
 \label{Aphiconst}
\end{eqnarray}

\noindent
These equations show clearly that nonzero holonomy exists on equatorial
orbits in the SD geometry.  We point out that the expressions are consistent
with the results for the two-sphere in flat space.  As $r \rightarrow \infty$,
the holonomy goes to zero.  However, for finite $r$,
a deficit angle exists after transport through and angular displacement of
$2\pi$ except when
$n(1-\frac{2M}r)^{1/2}$ is equal to an integer! It is this
``quantization" of holonomy that is initially striking.
Nevertheless, it is true and can be understood as follows:  Suppose that on
an orbit of radius $r_1$, the holonomy ``closes" and there is no deficit.
At another orbit $r_2$ slightly farther out, there must be a nonzero holonomy
because
the space between the two orbits is curved and according to the Gauss-Bonnet
theorem, the deficit angle is the integral of the Gaussian curvature over
the area enclosed by the two curves.
Had we considered only the holonomy intrinsic to the SD two-sphere,
the result would
have been the same as for the two-sphere in Euclidean space---zero.  The
difference is that we are here considering parallel transport in the full
space, which is the space relevant to local physics, and which is curved.
The quantization condition for holonomy invariance implies that

\begin{equation}
 r = \frac{2M}{1 - m^2/n^2},
\end{equation}

\noindent
with $m$ a non-zero integer.  Since we require $r > 2M$, we must have
$0<m^2/n^2 < 1$.  For fixed $m$, then, there is a minimum $n$ that
will give holonomy invariance.  In particular, no invariance
after $2\pi$ exists for $m = 1$.  After two loops invariance is
possible at $r = 8M/3$. In Table I we give examples of holonomy
invariance for various values of $m$ and $n$.

\begin{table}
\[
\begin{tabular}{|c||c|c|c|c|}\hline
 & n=1 & n=2 & n=3 & n=4 \\ \hline\hline
 m=1 & x & 2.667 & 2.25 & 2.133 \\ \hline
 m=2 & x & x & 3.6 & 2.667 \\ \hline
 m=3 & x & x & x & 6.4 \\ \hline
 m=4 & x & x & x & x \\ \hline
\end{tabular}
\]
\caption{\footnotesize{The table gives the radii $r$ (in units of $M$)
 at which holonomy invariance for constant-time circular orbits is achieved
 for integer $m$ and $n =$ number of circuits. An $x$ indicates
 that no invariance is possible for those values of $m$ and $n$.}}
\end{table}

\subsection{Timelike circles}
We consider now timelike circles.  The tetrad components of the
tangent vector are given by $X^{a} =
(\mu(1-2M/r)^{\frac{1}{2}},0,0,r)$.  For timelike curves we
require the squared magnitude to be less than zero, or $r^{3}+
2M\mu^2 - \mu^2r <0$. For such circles with radius $r > r_{crit}$,
we have $A^r(0) = \beta, \ A^t(0) = (\mu M/\omega r^2)\alpha +
\gamma$ and $A^\phi(0) = \omega^{-1}(1-2M/r)^{1/2}\alpha +
\delta$.  Hence from (\ref{Ar})-(\ref{Aphi}) after $n$ loops:

\begin{eqnarray}
\Delta A^r(2n\pi ) &=&\alpha \sin \omega 2n\pi +\beta (\cos \omega 2n\pi
-1),\\
\Delta A^t(2n\pi ) &=&\frac{\mu M}{\omega r^2}(\alpha (\cos \omega 2n\pi
-1)-\beta \sin \omega 2n\pi ), \\
\Delta A^\phi (2n\pi ) &=&\frac 1\omega
(1-\frac{2M}r)^{1/2}(\alpha (\cos \omega 2n\pi -1)-\beta \sin
\omega 2n\pi )\\
\Delta A_{\;}^\theta (2n\pi )&= & 0.\nonumber
\end{eqnarray}

\noindent
where $\Delta$ represents the difference in the components before and
after transport.  For invariance all the $\Delta$'s must vanish, which is
achieved when $\omega 2\pi n$ is an integer.  However, since $\omega$ depends
in a nontrivial way on $\mu^2M^2/r^4$, we will only examine the case of
greatest interest.

\subsection{Circular Geodesics}\label{Circgeo}

A geodesic is obtained by parallel transporting a tangent vector
along its integral curve. Note for the tangent vector above that
$X^r = 0$. Eqs. (\ref{DAt})-(\ref{DAphi}) show that for any vector
with $A^r = 0$, all components are constant with

\[
A^\phi =\mu\frac M{r^2}(1-\frac{2M}r)^{-1/2}\,A^t \nonumber.
\]
Applying this condition to $X$ itself yields
\[
\mu\frac M{r^2}(1-\frac{2M}r)^{-1/2}=X^\phi /X^t=r/((1-\frac{2M}r)^{1/2}\mu),
\]
which gives

\begin{equation}
\mu^2M=r^3 .
\end{equation}

\noindent
This is Kepler's third law for relativistic orbits.
Now, the magnitude of $X$ is $3(1-r/3M)r^2$.  Thus a circular
geodesic is timelike for $r > 3M$, spacelike for $r < 3M$ and null for
$ r = 3M$, as required (this is the last stable orbit for photons).
Furthermore, we see that for circular geodesics
the Kepler condition implies $\omega^2 = 1 - 3M/r$, which means that $r_{crit} =
3M$.  In the context of the so-called optical geometry (see \cite{AL, SM} and
references therein), this surprising result is perhaps not so surprising. In
the optical metric (where the usual SD spatial metric is multiplied by
$(1-2M/r)^{-1}$, a conformal rescaling), {\em all} dynamical effects of
circular motion reverse at $r = 3M$, for example the direction of
centrifugal force and the direction of precession of
gyroscopes.  To this list we may add another effect: the change of parallely
propagated solutions along circular geodesics from oscillatory to exponential.

\begin{figure}[htb]
\vbox{\hfil\scalebox{0.4}
  {\includegraphics{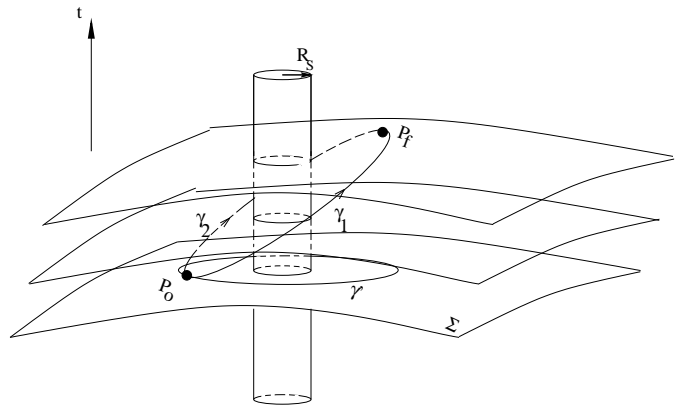}}\hfil}
  {\caption{\footnotesize{Circular geodesics in Schwarzschild spacetime. $P_0$
  is the initial point, $P_f$ is the final point; $\Sigma$ represents
  the initial hypersuface and $\gamma_1$ and
  $\gamma_2$ circular geodesics.}}}
\end{figure}

The expression for $\omega$ allows us
to compute holonomy invariance for geodesic paths whose spatial
projections are circular orbits (Fig. 1).  In this case,
invariance is achieved when

\begin{equation}
(1-\frac{3M}r)^{1/2}n = m.
\end{equation}

\noindent
Clearly the condition that $r > 2M$ puts
restrictions on the allowed values of $n$ that give holonomy
invariance.  In particular, no holonomy invariance exists for
$n=1$; for $n=2$, $r = 12/3M$ gives invariance after transport
through an angle $2\pi$. Table II gives other values of holonomy
for circular geodesics.

\begin{table}
\[
\begin{tabular}{|c||c|c|c|c|}\hline
        & n=1 & n=2 & n=3 & n=4 \\ \hline\hline
    m=1 & x & 4 & 3.375 & 3.2 \\ \hline
    m=2 & x & x & 5.4 & 4 \\ \hline
    m=3 & x & x & x & 9.6 \\ \hline
    m=4 & x & x & x & x \\ \hline
\end{tabular}
\]
\caption{\footnotesize{The table gives the radii $r$ (in units of $M$)
at which holonomy invariance for circular geodesics is achieved for integer
$m$ and $n =$ number of circuits. As in Table I. an $x$ indicates that no
invariance is possible for those values of $m$ and $n$.}}
\end{table}

%-----------------------------------------------------------------------------

\section{Radial Holonomy}
\setcounter{equation}{0} \label{radial}
%-----------------------------------------------------------------------------

In the case of radial paths, we take the tangent vector to be $X^\mu = [X^t,
X^r,0,0]$.  We may choose $t$ or $r$ to be the curve parameter $\lambda$,
and thus may set either $X^t$ or $X^r$ to 1.  However, in this case one
should avoid putting $dt/dr = constant$, or vice versa, since this will
in general not be true for radial geodesics. Let us now specialize to
radial null geodesics. To find them, note
that the tetrad components of the tangent vector are
\begin{equation}
X^a={\Bigl(}(1-\frac{2M}r)^{1/2}X^t,(1-\frac{2M}r)^{-1/2}X^r,0,0{\Bigr)},
\end{equation}
with magnitude
\[
X^2=-(1-\frac{2M}r)(X^t)^2+(1-\frac{2M}r)^{-1}(X^r)^2.
\]
This vector is null if and only if
$(1-\frac{2M}r)(X^t)^2=(1-\frac{2M}r)^{-1}(X^r)^2$, which implies
\begin{equation}
 (X^t)^2=(1-\frac{2M}r)^{-2}(X^r)^2.
\end{equation}
With a curve parameter $r$, eqs.(\ref{TransComp}) reduce to two nontrivial
equations
\begin{eqnarray}
 \frac{dA^t}{dr} + \frac{M}{r^2}(1-\frac{2M}{r})^{-1}A^r & = & 0, \\
 \frac{dA^r}{dr} + \frac{M}{r^2}(1-\frac{2M}{r})^{-1}A^t & = & 0,
\end{eqnarray}
which are easily integrated to give the general solution
\begin{eqnarray}
 A^{t} &=& c_{1}(1-2M/r)^{-\frac{1}{2}} + c_{2}(1-2M/r)^{\frac{1}{2}},\\
 A^{r} &=& c_{1}(1-2M/r)^{-\frac{1}{2}} - c_{2}(1-2M/r)^{\frac{1}{2}}.
\end{eqnarray}
The integration constants $c_{1}$ and $c_{2}$ may be fixed as $1$ and $0$
respectively by imposing that
$A^{t}=X^{t}$ and $A^{r} = X^{r}$. This yields
\begin{equation}
 A^{t}=A^{r}=(1-2M/r)^{-\frac{1}{2}}.
\end{equation}

\noindent
Turning to constant $r,\theta,\phi$ curves, we now have $X^r = 0$ and  $t$
may be taken as the curve parameter. The solutions
\begin{eqnarray}
A^t(t) &=& A^t(0)\cosh (\frac M{r^2}t)-A^r(0)\sinh (\frac
M{r^2}t),\;\nonumber\\
 A^r(t) &=& A^r(0)\cosh (\frac
M{r^2}t)-A^t(0)\sinh (\frac M{r^2}t), \label{constr}
\end{eqnarray}
with $A^{\theta}$ and $A^{\phi}$ constant are easily found from
eqs.(\ref{TransComp}) by setting $d\phi = 0$ and integrating.
These expressions may then be used to construct holonomy along sample paths.
For example, consider a constant $r,\phi$ curve as above.  We transport a vector
$A = (A^{t}(0),A^{r}(0),A^{\theta}(0),A^{\phi}(0))$ along a radial
null geodesic from a point A at a radius $r_o$ inward to
a point B at $r = r_1$, then outward along a radial null geodesic to $r = r_0$
again to a point C, as in Fig. 2. The radial contributions cancel and
we are left with the holonomic change

\begin{eqnarray}
  \Delta A^{t} &=& A^{t}(0){\Bigl(}\cosh(\frac{M}{r_o^2}t_{C}) - 1{\Bigr)} -
  A^{r}(0)\sinh(\frac{M}{r_o^2}t_{C}),\nonumber\\
  \Delta A^{r} &=& A^{r}(0){\Bigl(}\cosh(\frac{M}{r_o^2}t_{C}) - 1{\Bigr)},
\end{eqnarray}
where we have taken $t_{A} = 0$. Note that as $r_{0}\rightarrow\infty$ the
change in the vector due to parallel transport vanishes. Indeed this is to be
expected from the asymptotic flatness of the Schwarzschild spacetime.
This result can be found in \cite{Bollini}. However, we emphasize that
the $r, \phi = constant$ path is far from generic; any
spaceship on such an orbit would have to fire rockets to remain in position
and thus this is definitely not geodesic motion. A more ``realistic" path
would be a wedge, bounded by two circular orbits at radii $r_0$ and
$r_1$ and two radial geodesics, as in Fig. 3.

\begin{figure}[htb]
\vbox{\hfil\scalebox{0.5}
  {\includegraphics{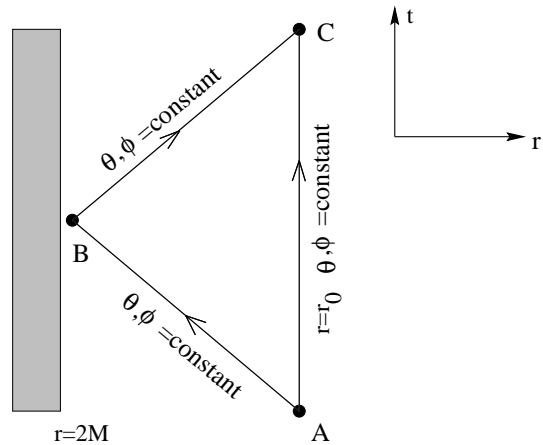}}\hfil}
  \caption{\footnotesize{Spacetime diagram for a path traversed radially inward from
  a point $A$ at some finite radius $r=r_{0}$ to a point $B$ neighborhood of
  the Schwarzschild radius and then radially outward to $C$, at the radius
  $r_0$. We assume that the radial paths, $AB$ and $BC$ are radial ingoing and
  outgoing geodesics respectively.}}
\end{figure}

\begin{figure}[htb]
\vbox{\hfil\scalebox{0.4}
  {\includegraphics{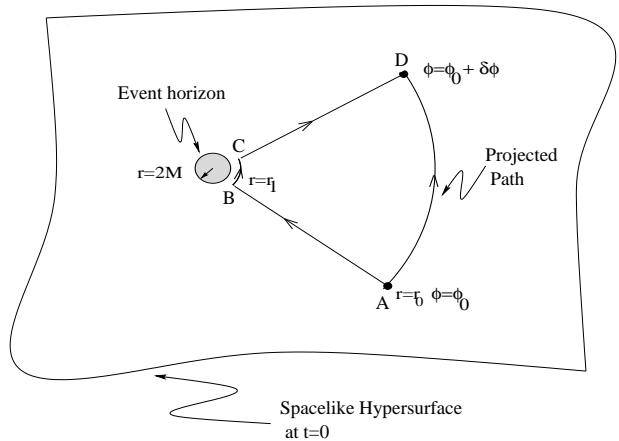}}\hfil}
  \caption{\footnotesize{Projection of a `wedge shaped' path onto the initial spatial
  hypersurface.}}
\end{figure}

Again, the radial contributions cancel and if one compares the
vector transported directly from A to D with one transported
around the loop from A to B to C to D, one gets holonomy for, say,
the $A^r$ component (setting $\alpha = 0$ which yields $\beta =
A^{r}(0)$)
\begin{equation}
 \Delta A^r = A^r(0)\left\{\cos(\omega_0\phi)
  - \cos[(1-\frac{3M}{r_1})^{1/2}\phi]\right\},
  \label{DeltaAr}
\end{equation}
if $r_0, r_1 > r_{crit}$. We have assumed for illustration that
one circular orbit is a geodesic but not the other because, in general.
both cannot be geodesic.  That is, although the angles are the same, the time
taken to traverse the outer orbit is exponentially longer than the
time taken to traverse the inner orbit.  This in turn is because the time
taken for photons to traverse the radial null geodesics is given by
\[
t = |r_*| = \int \frac{dr}{(1 - \frac{2M}{r})^{1/2}}.
\]
Consequently, $t_0$ is not independent of $t_1$.  However, for geodesics,
Kepler's law determines the time taken to traverse a given angle; thus
when both orbits are geodesics the system is over-determined. With
$\omega$ given by (\ref{varpi}), equating the arguments in (\ref{DeltaAr})
gives for holonomy invariance modulo $2\pi$
\[
r_1 = \frac{3r_0^4}{2r_0^3 + \mu^2M}.
\]
If $\mu^2M$ is small compared to $r_0^3$, this yields $r_1 = 3/2 r_0$. Note
that if $r_0$ is taken to be a geodesic orbit, then $\mu^2M = r_0^3$ and
$r_1 = r_0$, showing that in this limit no holonomy invariance is possible
modulo $2\pi$ unless the two geodesic orbits are the same.  For higher number
of circuits invariance is possible.\\

In Section \ref{Discussion} we will discuss possible applications of holonomy.
Before doing this, however, it will be helpful
 to motivate the discussion by recovering some of the previous results
 via the loop formalism and for spinors, and investigating a few properties
 of holonomy in the Reissner-Nordstr\"om background.

%-----------------------------------------------------------------------------
\section{Loop formulation}
\setcounter{equation}{0} \label{sec}
%-----------------------------------------------------------------------------

As is clear from the previous discussion, the result of parallel transport is
path dependent. This may be stated slightly more rigorously as follows: Let
${\cal M}$
be some $n$-dimensional manifold (which we shall choose as our Schwarzschild
background). For
each closed curve $\gamma: [0,1]\rightarrow {\cal M}$ with base point $p$, parallel
transport associates some $GL(n,{\bf R})$-valued operator $U(\gamma)$ acting on
$T_{p}$ called the {\it holonomy}. The concept of holonomy has
appeared in many guises in physics from
lattice gauge theories {\cite{Wilson}} to loop quantum gravity
{\cite{Gambini-Pullin1996}} (see also \cite{Bollini, Bezerra96,Bezerra87} for
further details) with just as many aliases. Thus it is known, for example, as the
{\it Wu-Yang phase factor} in particle physics. The holonomy associated with the
parallel transport around $\gamma$ is a linear map from the
tangent space at a point $p\in {\cal M}$ into itself, realized by the path-ordered
exponential

\begin{equation}
 U(\gamma) = P\,e^{-\int_{\gamma}\,\Gamma_{\mu}dx^{\mu}},
\end{equation}
where $\Gamma_{\mu}$ is the tetradic connection on ${\cal
M}$.\footnote{We thank Goh Liang Zhen for pointing out a sign
error in several works on this subject including \cite{Bezerra87}
and \cite{Bollini} and that the correct sign in the exponent is
negative.} Choosing ${\cal M}$ to be a $4$-dimensional Lorentzian
manifold with Schwarzschild metric we get,
\begin{equation}
\Gamma_\mu = \left(
\begin{array}{llll}
0 & \Gamma _{\;rt}^t & 0 & 0 \\
\Gamma _{\;tt}^r & 0 &  \Gamma _{\;\theta\theta}^r & \Gamma _{\;\phi\phi}^r \\
0 & \Gamma _{\;r\theta}^\theta & 0 & \Gamma^{\theta}_{\phi\phi} \\
0 &  \Gamma _{\;r\phi}^\phi & \Gamma^{\phi}_{\theta\phi} & 0
\end{array}
\right)
\label{Gamma}
\end{equation}
where the connection coefficients are as derived from (\ref{connectionForms}).\\

To illustrate an explicit representation of $U(\gamma)$ consider
circular orbits, $\Gamma_\mu dx^\mu = \Gamma_t dt + \Gamma_\phi
d\phi$. Clearly the matrix elements $\Gamma _{\;\theta\theta}^r$
and $\Gamma _{\;r\theta}^\theta$ of ({\ref{Gamma}}) vanish.
With $\mu \equiv dt/d\phi$ as before, we can write
$\int \Gamma_t dt + \Gamma_\phi d\phi =
2\pi[\Gamma_t \mu + \Gamma_\phi] \equiv 2\pi\Gamma$.
We then need to evaluate
\begin{equation}
U = e^{-\int\Gamma_\mu dx^\mu} = e^{-2\pi\Gamma}
        = 1 - 2\pi\Gamma + \frac{(2\pi\Gamma)^2}{2!}
            - \frac{(2\pi\Gamma)^3}{3!} + ....
\label{UTaylor_series}
\end{equation}
A little algebra shows that $\Gamma^3 = -S^2\Gamma$, where $S^2
\equiv (\Gamma_{\;r\phi}^{\phi})^2 - \mu^2(\Gamma _{\;rt}^t)^2$ in
terms of which the Taylor series (\ref{UTaylor_series}) becomes

\begin{equation}
U = 1 - \frac{\Gamma}{S}\sin(2\pi S) +
                    \frac{\Gamma^2}{S^2}(1-\cos(2\pi S)). \label{Ucirc}
\end{equation}
We see, somewhat unexpectedly, that $S^2 = 1-\frac{2M}{r}-\frac{M^2\mu^2}
{r^4}= \omega^2$. The final vector ${\bf A_f}$ after parallel transport is
given merely by
${\bf U \cdot A_o }= U^\alpha\;_\beta A_o^\beta$.  Thus, the deficit angle between
the initial and final vector is simply ${\bf A_o \cdot A_f = A_o \cdot U \cdot
A_o}= |A_o||A_f|\cos\chi$, where $\chi$ is the deficit angle.  Thus, in
general, ${\bf \hat A_o
\cdot \hat A_f } = \cos\chi$ where ${\bf\hat A_o}$ and ${\bf\hat A_f}$
are now unit vectors\footnote{Here we are using ordinary matrix notation, but
note that ${\bf A \cdot B }\equiv \eta_{\mu\nu} A^\mu B^\nu$.  Also note that
U preserves norms and hence is an orthogonal transformation.}.   Given $U$ as in
(\ref{Ucirc})
we find ${\bf \hat A_o \cdot \hat A_f } = \cos(2\pi S)$.  For constant-
time circles, $\mu = 0$, which immediately gives
 ${\bf \hat A_o \cdot \hat A_f } = \cos(2\pi(1-2M/r)^{1/2})$, in agreement with
 what is obtained from (\ref{Arconst}) and (\ref{Aphiconst}). This coincides
 with the corresponding result in \cite{Bollini}. If we consider the
 constant $r,\phi$ orbits of the previous section,
  then the matrix $\Gamma$
 contains only the $r,t$ and $t,r$ components ($= M/r^2$).  Computing $U$
 in an analogous way as above yields
 \begin{equation}
 U =  \left(
 \begin{array}{llll}
 \cosh(\frac{Mt}{r^2}) & -\sinh(\frac{Mt}{r^2}) & 0 & 0 \vspace{1mm}\\
 -\sinh(\frac{Mt}{r^2}) & \cosh(\frac{Mt}{r^2}) & 0 & 0 \vspace{1mm}\\
 0                    & 0                    & 1 & 0\\
 0                    & 0                    & 0&  1
 \end{array}
 \right)
 \end{equation}
Taking the dot product gives for the deficit angle
\begin{equation}
 \cos\chi = \frac{1}{|A|^2}
        [(-(A^t)^2 + (A^r)^2)\cosh(Mt/r^2) + (A^\theta)^2 + (A^\phi)^2],
\end{equation}
where $|A|^2 = -(A^t)^2 + (A^r)^2 + (A^\theta)^2 + (A^\phi)^2$, in
agreement with what is obtained from (\ref{constr}). (Note,
however, that $\chi$ will not always be real.)

%-----------------------------------------------------------------------------
\section{Spinors}
\setcounter{equation}{0} \label{Spinors}
%-----------------------------------------------------------------------------
The vectors we have been parallel transporting are ordinary
vectors in spacetime; they could as well be genuine arrows or
gyroscope axes. Apart from the behavior of gyroscopes, the more
relevant question for physics is, How do particle wave-functions
behave under parallel transport? Quantum field theory in curved
space is almost invariably the study of $(0,0)$ representations of
the Lorentz group i.e., spin-0 fields.  While clearly a vast
simplification, the theory is still rich enough to reveal such
semi-classical artifacts as Hawking radiation. Nevertheless,
unless we have a full treatment of higher rank representations,
the theory cannot be considered complete. From the point of view
of holonomy, spinor , vector and tensor fields are obviously
considerably more interesting than scalar fields. How then does
parallel transport affect such higher spin fields? In this section
we describe the parallel transport of spinors on the Schwarzschild
manifold \footnote{Particle wave-functions live in an
appropriately constructed Hilbert bundle over spacetime but are
affected by spacetime transport. Here we describe the parallel
transport of {\it classical spinors} in spacetime assuming that
the formalism will apply to any theory combining quantum and
spacetime transport; see Section \ref{Discussion}.}. We begin by
briefly reviewing some necessary results. Fuller treatments of
spinor formalism can be found in \cite{Stewart},\cite{Bade53} and
\cite{Pirani}.  The results are essentially the same as the
previous but with the expected factor of $1/2$ in the relevant
arguments.

One defines the spinor covariant derivative in the same way as the tensorial
covariant derivative:
\begin{equation}
\nabla _\mu\kappa _B=\partial _\mu\kappa _B-\Gamma _{\;B\mu}^C\kappa_C .
\label{spincov}
\end{equation}
Here, however, lower-case Greek indices ($\mu$) with range 0,1,2,3,
will represent coordinates; lower-case Latin indices with the same
range will represent tetrad components and upper-case Latin indices (B,C),
with range 0,1 will
represent spinor components (for two-component spinors).
 The spinor connection is given in terms of
the tetrad rotation coefficients $\Gamma _{\;b\mu}^d$ by

\begin{equation}
\Gamma _{\;B\mu}^C=\frac{1}{12}\sigma _d^{\;C\dot{Y}}\left( \sigma _{\;B\dot{Y}
}^b\Gamma _{\;b\mu}^d+\partial _\mu\sigma _{\;B\dot{Y}}^b\right) .
\label{spingamma}
\end{equation}
The matrices $\sigma _d^{\;C\dot{Y}}$ can be thought of as extended set of
Pauli matrices, which associate every vector with a second-rank Hermitian
spinor, i.e., $A_{\dot{L}Y} = \sigma^k_{\dot{L}Y}A_k$. The dot over an
index is conventionally used to indicate complex conjugation: $A_{\dot{L}Y}$
is the complex conjugate of $A_{Y\dot{L}}$.  However, for our purposes
a dotted index is to be treated like any other.  A set of basis spinors
can be chosen as
\begin{eqnarray*}
\sigma _0^{\;B\dot{X}}= \frac {1}{\sqrt{2}}\left(
\begin{array}{cc}
1 & 0 \\
0 & 1
\end{array}
\right) \ ; \
\sigma _1^{\;B\dot{X}} &=&\frac {1}{\sqrt{2}}\left(
\begin{array}{cc}
0 & 1 \\
1 & 0
\end{array}
\right) \\
\sigma _2^{\;B\dot{X}}=\frac {1}{\sqrt{2}}\left(
\begin{array}{cc}
0 & -i \\
i & 0
\end{array}
\right) \ ; \
\sigma _3^{\;B\dot{X}} &=&\frac {1}{\sqrt{2}}\left(
\begin{array}{cc}
1 & 0 \\
0 & -1
\end{array}
\right)
\end{eqnarray*}
To construct the spinor connections we need the $\sigma$'s with lowered indices,
which are defined by the identity
\begin{equation}
\sigma _a^{\;B\dot{Y}}\sigma _{\;B\dot{Y}}^b=\delta _a^b ,
\end{equation}
or equivalently,
\begin{equation}
\sigma _b^{\;C\dot{X}}\sigma _{\;B\dot{Y}}^b=\delta _B^C\delta _{\dot{Y}}^{%
\dot{X}}\;
\label{ddelta}.
\end{equation}
One often terms $\sigma _{\;B\dot{Y}}^b$  the inverse of $\sigma
_b^{\;C\dot{X}}$, but
they are {\em not} multiplicative inverses.  To construct $\sigma
_a^{\;B\dot{Y}}$ we lower indices with the fundamental spinor,
 $\epsilon_{CB}$, which can be chosen as the Levi-Civita permutation operator:
\begin{equation}
\epsilon_{CB} =
\left(
\begin{array}{ll}
0 & 1 \\
-1 & 0
\end{array}
\right) = \epsilon^{CB}.
\end{equation}
Because the permutation operator is antisymmetric, when manipulating spinor
indices it is crucial that the indices to be
lowered are aligned with the corresponding indices of the
permutation symbol:
\begin{eqnarray}
\sigma^b_{B\dot{X}} &=& \eta^{bd}\;\sigma_d^{C\dot{Y}}\epsilon_{CB}\;
                \epsilon_{\dot{Y}\dot{X}}\\
             &=&-\eta^{bd}\;\epsilon_{BC}\;\sigma_d^{C\dot{Y}}\;
                    \epsilon_{\dot{Y}\dot{X}}.
\label{siginv}
\end{eqnarray}
Because we have adopted tetradic connections, the Minkowski metric
$\eta^{bd}$ is
used to raise and lower the tetrad indices; otherwise
the metric tensor $g^{\mu\nu}$ would be substituted. Eq (\ref{siginv}) is
perhaps more clearly written as
\begin{equation}
\sigma^b_{down} = -\eta^{bd} \epsilon^T \sigma_d^{up}\epsilon ,
\end{equation}
Where $\epsilon^T$ is the transpose of $\epsilon$.  With this equation we find
\begin{eqnarray*}
\sigma _{\;B\dot{X}}^0 =\frac {1}{\sqrt{2}}\left(
\begin{array}{cc}
1 & 0 \\
0 & 1
\end{array}
\right) \ ; \
\sigma _{\;B\dot{X}}^1 &=&\frac {1}{\sqrt{2}}\left(
\begin{array}{cc}
0 & 1 \\
1 & 0
\end{array}
\right) \\
\sigma _{\;B\dot{X}}^2 =\frac {1}{\sqrt{2}}\left(
\begin{array}{cc}
0 & i \\
-i & 0
\end{array}
\right) \ ; \
\sigma _{\;B\dot{X}}^3 &=&\frac {1}{\sqrt{2}}\left(
\begin{array}{cc}
1 & 0 \\
0 & -1
\end{array}
\right)
\end{eqnarray*}
Clearly $\sigma _{\;B\dot{X}}^2$ is not the multiplicative inverse of
$\sigma _2^{\;B\dot{X}}$. One easily verifies that the $\sigma$ matrices satisfy
identity
(\ref{ddelta}).  This formalism is all that is required to compute
the holonomy associated
with various paths in the SD geometry.\\

For circular orbits of constant time, the spin covariant derivative
(\ref{spincov}) is
\begin{equation}
\nabla _\phi \kappa _B=\partial _\phi \kappa _B-\Gamma _{\;B\phi }^C\kappa
_C ,
\label{spingammcirc}
\end{equation}
where
\begin{equation}
\Gamma _{\;B\phi }^C=\frac {1}{2}\sigma _d^{\;C\dot{Y}}\left( \sigma _{\;B\dot{Y%
}}^b\Gamma _{\;b\phi }^d+\partial _\phi \sigma _{\;B\dot{Y}}^b\right)
=\frac {1}{2}\sigma _d^{\;C\dot{Y}}\sigma _{\;B\dot{Y}}^b\Gamma _{\;b\phi }^d.
\end{equation}
The second equality follows from the fact that we have chosen the
$\sigma$'s to be constant.  The tetradic connections are exactly those
derived from (\ref{connectionForms}) (with $Q=0$) and so
\begin{eqnarray}
\Gamma _{\;B\phi }^C &=&\frac {1}{2}\left( \sigma _3^{\;C\dot{Y}}\sigma _{\;B%
\dot{Y}}^1\Gamma _{\;r\phi }^\phi +\sigma _1^{\;C\dot{Y}}\sigma _{\;B\dot{Y}%
}^3\Gamma _{\;\phi \phi }^r\right) \nonumber\\
&=&\frac {1}{2}(1-\frac{2M}r)^{1/2}\left( \sigma _3^{\;C\dot{Y}}\sigma _{\;B%
\dot{Y}}^1-\sigma _1^{\;C\dot{Y}}\sigma _{\;B\dot{Y}}^3\right) .
\label{spingammphi}
\end{eqnarray}
Working this out with the above $\sigma$-matrices gives two nonzero spin
connections:
\begin{equation}
\Gamma _{\;1\phi }^0 = \frac{1}{2}(1-\frac{2M}r)^{1/2} = - \Gamma _{\;0t}^1.
\end{equation}
These can summarized by the formula
\begin{equation}
\Gamma _{\;Bt}^C=\frac{1}{2}\frac M{r^2}\left( \delta _0^C\delta _B^1+\delta
_1^C\delta _B^0\right).
\end{equation}
Note that these spinor connections are, as expected, exactly one-half of
the corresponding tetrad connections.  Not surprisingly, parallel transport
produces holonomy invariance after a circuit of $4\pi$ rather than $2\pi$.
We have from (\ref{spingammcirc}) for the two components of the spinor
(B = 0,1):
\begin{equation}
\partial _\phi \kappa _0+\frac{1}{2}(1-\frac{2M}r)^{1/2}\kappa _1 = 0.
\end{equation}
and
\begin{equation}
\partial _\phi \kappa _1-\frac {1}{2}(1-\frac{2M}r)^{1/2}\kappa _0 = 0.
\end{equation}
As before we can solve these by differentiating the first and substituting in
the second, which gives
\begin{equation}
\partial _\phi ^2\kappa _0=-\frac {1}{4}(1-\frac{2M}r)\kappa _0.
\end{equation}
Thus, in analogy with (\ref{Ar}), these equations have the solutions
\begin{eqnarray}
\kappa_0 &=& \kappa_0(0)\cos(\omega\phi) - \kappa_1(0)\sin(\omega\phi), \nonumber\\
\kappa_1 &=& \kappa_0(0)\sin(\omega\phi) + \kappa_1(0)\cos(\omega\phi).
\end{eqnarray}
where $\omega = \frac{1}{2}(1-\frac{2M}{r})^{1/2}$ exhibits the required
factor of $\frac{1}{2}$.\\

For general circular orbits we require,
$X^\mu\nabla_\mu\kappa_B = 0 $.  Taking $X^\mu = [\mu,0,0,1]$, with
$\mu = dt/d\phi$
as before leads to the two equations (note signs):
\begin{eqnarray}
\frac{d\kappa_0}{d\phi}  -\frac{1}{2}(\frac{\mu M}{r^2}
                - (1-\frac{2M}{r})^{1/2})\kappa_1 &=& 0,
                \nonumber\\
\frac{d\kappa_1}{d\phi}  -\frac{1}{2}(\frac{\mu M}{r^2}
                + (1-\frac{2M}{r})^{1/2})\kappa_0 & = & 0.
\end{eqnarray}
Differentiating the first and substituting in the second gives
\begin{equation}
\frac{d^2\kappa_0}{d\phi^2}
        +\frac{1}{4}\left(1-\frac{2M}{r}-\frac{\mu^2M^2}{r^4}\right)\kappa_0
                = 0.
\end{equation}
Apart from a factor of $1/2$ in the second term, this is similar to the
corresponding equation for vector transport and hence, apart from a rescaled
frequency, its solutions exhibit the same qualitative behavior as (\ref{Ar}).
The remaining properties of circular orbits
already discussed for the vectors persist for the spinor case.
Radial holonomy can be worked out in similar fashion. In analogy
to (\ref{spingammphi}) we have
\begin{equation}
\Gamma _{\;Bt}^C = \frac{1}{2}\left( \sigma _0^{\;C\dot{Y}}\sigma _{\;B\dot{Y}%
}^1\Gamma _{\;rt}^t+\sigma _1^{\;C\dot{Y}}\sigma _{\;B\dot{Y}}^0\Gamma
_{\;tt}^r\right),
\label{spincont}
\end{equation}
which gives two nonzero spin connections
\begin{equation}
\Gamma _{\;1t}^0 =\frac {1}{2}\frac M{r^2} = \Gamma _{\;0t}^1
\end{equation}
Thus, for the $r, \phi$ = constant paths, the parallel transport
condition $\nabla _t\kappa _B=\partial _t\kappa _B-\Gamma _{\;Bt}^C\kappa _C=0$
gives for $B = 0$
\begin{equation}
\partial _t\kappa _0-\frac{1}{2}\frac M{r^2}\kappa _1 = 0,
\end{equation}
and for $B = 1$
\begin{equation}
\partial _t\kappa _1-\frac{1}{2}\frac M{r^2}\kappa _0 = 0.
\end{equation}
This pair of equation, then, leads to the exponential solution corresponding
to (\ref{constr}):
\begin{eqnarray}
\kappa_0(t) &=& \kappa_0(0)\cosh (\frac {Mt}{2r^2})-
        \kappa_1(0)\sinh (\frac {Mt}{2r^2}),\;\nonumber\\
\kappa_1(t) &=& \kappa_1(0)\cosh (\frac {Mt}{2r^2})-\kappa_0(0)(\sinh \frac
            {Mt}{2r^2}),
\label{constrspin}
\end{eqnarray}
For radial paths we need $\Gamma^C_{Br} \propto \Gamma^D_{br}$.
However all tetradic connections with $r$ in the last place are zero.  Thus
along radial paths, $\nabla_r\kappa_B = \partial\kappa_B$ and for
tangent vector $X^\mu$, the parallel
transport condition becomes $X^t\nabla_t\kappa_B + \partial_r\kappa_B = 0$.
Specializing to null geodesics as before leads to
\begin{equation}
\frac{d^2\kappa_0}{dr^2} +\frac{2}{r}\left[1 + \frac{M}{(r-2M)}\right]
            \frac{d\kappa_0}{dr}
            - \frac{M^2}{4r^4}(1-\frac{2m}{r})^{-2}\kappa_0 = 0,
            \label{D2kappa}
\end{equation}
which has solution
\begin{equation}
\kappa_0 = c_1\frac{r^{1/4}}{(r - 2M)^{1/4}} + c_2\frac{(r-2M)^{1/4}}{r^{1/4}},
\end{equation}
We also note that the holonomy map $U$ will be unchanged except for a
factor of $1/2$ before the elements; the rows and columns should be
relabeled $(0,1,\dot{0},\dot{1})$. Thus spinor parallel transport in spacetime
is not qualitatively dissimilar to that of vector transport and certainly does
not manifest further surprises beyond those recognized in the vector case.

%-----------------------------------------------------------------------------
\section{Holonomy in the Reissner-Nordstr\"om geometry}
\setcounter{equation}{0} \label{RN}
%-----------------------------------------------------------------------------

For reasons to be discussed below, it is of some interest to compare features
of holonomy in the Reissner-Nordstr\"om background with the results already
obtained for SD.  The calculations are identical except we now use the full
connections (\ref{connectionForms}) with $Q \ne 0$.  We restrict our comments to
circular orbits.  The differential equation for $A^r$ (\ref{DAr}) becomes
\begin{equation}
\frac{ dA_{\;\ }^r}{d\phi} + \left(\frac M{r^2}-\frac{Q^2}{r^3}\right)A^t\mu
        -\left(1-\frac{2M}r + \frac{Q^2}{r^2}\right)^{1/2}A^\phi =0,
\label{RNDAr}\\
\end{equation}
with the same oscillatory and exponential solutions (\ref{Ar}) except
 that now
\begin{equation}
\omega^2 = 1 - \frac{2M}{r} + \frac{Q^2}{r^2} - \mu^2\left(\frac{M}{r^2}
        -\frac{Q^2}{r^3}\right)^2.
        \label{RNfreq}
\end{equation}
The tetrad components of the tangent vector for RN are
\begin{eqnarray}
 X^{a} = {\Bigl(}
           (1-{\frac{2M}{r}} + {\frac{Q^2}{r^2}})^{1/2}\mu,0,0,r
     {\Bigr)}.
 \label{RNtetrad}
\end{eqnarray}
Using the geodesic condition $A^r = 0$ in (\ref{RNDAr}) and applying it to
the tetrad components, as we did in Section \ref{Circgeo}, yields for
Kepler's third law
\begin{equation}
r^3 = \mu^2\left(M - \frac{Q^2}{r}\right).
\end{equation}
With this condition, (\ref{RNfreq}) becomes
\begin{equation}
\omega^2 = 1 - \frac{3M}{r} + \frac{2Q^2}{r^2},
\end{equation}
for circular geodesics.  Note that in distinction to the SD case we now have
a quadratic equation for $\omega$, which will in general have two roots for
$r_{crit}$.  They are
\begin{equation}
r_{crit\pm} = \frac{3M}{2}\left(1\pm \sqrt{1 - \frac{8Q^2}{9M^2}}\right).
\end{equation}
Now, the horizon in the RN solution is located at $r_+ \equiv M +
\sqrt{M^2 - Q^2}$. In general, $r_{crit-} < r_+$ and so the
negative root can be discarded. However, at extremality ($Q = M$)
the horizon is located at $r_+ = r_- = M$ and so we have two
physically relevant roots, $r_{crit-} = M$ and $r_{crit+} = 2M$.
Hence, null rotations around extremal RN black holes take place
both on the horizon and at $r = 2M$, in distinction  to the SD
case, where there is only one $r_{crit}$ (= 3M) and the solutions
on the horizon are oscillatory.  This surprising result is another
in the growing list of features that suggest extremal black holes
are of a qualitatively different class than their nonextremal
counterparts (see, e.g., \cite{LRS00}).

%-----------------------------------------------------------------------------
\section{Discussion}
\setcounter{equation}{0} \label{Discussion}
%-----------------------------------------------------------------------------

This investigation of parallel transport of vectors and spinors in the
SD geometry has revealed several surprising results. An important question,
though, is whether they are mere curiosities or whether they can be
linked to other interesting phenomena. Clearly
the testing of any of the results requires the curves along which the transport
is carried out to be timelike or at worst null, ruling out many
of the classes of curves that we have studied here. On the other hand, while
at first glance our study of
parallel transport along constant time curves may seem physically irrelevant, it
does serve as a simple toy scenario against which to test our calculations and
to compare with timelike and null results.\\

One important example of possible relevance is the
Einstein-Podolsky-Rosen (EPR) gedanken-experiment.  As is widely
known, when the participating particles move in a region of
non-negligible gravitational field, the standard approach to the
EPR paradox must be radically reformulated. This is as a direct
result of the fact that, in a curved spacetime even if the system
is prepared with the $z$-axes of the two particles aligned, this
in no way guarantees that after propagation in a gravitational
field their axes will remain aligned at the point of measurement.
It has recently been shown {\cite{Mensky2000}} that quantum
correlation between the spins of EPR particles in a gravitational
field may be formulated with the aid of parallel transport. As
such, some of the curves we have studied here may prove useful in
a proper analysis of EPR experiments in curved spaces.\\

It is also worth pointing out that in the holonomy band structure
discussed in Section \ref{Circular} we have an
 example of quantization that does
not depend on Planck's constant.  Can it be measured?  Gyroscopes
obey Fermi-Walker transport, in which the timelike vector of the
local tetrad is is held parallel to the tangent vector of the
curve.  In this way there are no spatial rotations.  Along
geodesics Fermi-Walker transport becomes parallel transport.
Nevertheless, given that the holonomy depends on $M/r \sim
10^{-8}$ for Earth, the experiment does not seem terribly
feasible (witness the difficulties attending the Stanford
gyroscope experiment). However, in the geometric optics approximation,
the polarization vector of light is parallel transported.  In
principle, then, one could measure polarization bands around
black holes, where $M/r \sim 1$.  In principle, one could also
measure electron interference, since the electron spin axis is
also parallel transported.\\

The holonomy properties we have investigated may be more interesting in
a quantum setting.
As mentioned earlier, the holonomy map $U$ was originally introduced
in gauge theories in the form
\begin{equation}
 \Phi(C) = P e^{ ig\oint A_\mu(x)dx^\mu}.
\end{equation}

\noindent
In this equation the $A_\mu$'s represent gauge potentials and, in
the case of electromagnetism, $\Phi$ represents the Dirac
phase factor, the observable in the Aharonov-Bohm experiment. In
the gravitational case, therefore, the $\Gamma$'s play the role
of the $A_\mu$. However, the analogy is not exact in the sense
that the holonomy in gauge theories take place in the internal
gauge space of the theory, whereas the ``phase" changes we have
described in this paper take place in real spacetime. Many
authors \cite{Asorey82,Coleman96,Sardan00,Bozo00,Dand89}have also
noticed the similarity between the holonomy map $U$ and the
expression for the phase that appears in quantum-mechanical path
integration: $exp(i \int L dt)$, where $L$ is the classical
action for a particle.

Here, however, although the path
integration does take place over space, the canonical position
and momenta that figure in the action form operators in Hilbert
space, not in spacetime. Nevertheless, it is reasonable to ask
whether the similarities among the three expressions are
coincidental or whether one can truly regard quantum evolution as
an example of parallel transport.  Specifically, what is the
relation, if any, between parallel transport in spacetime and
parallel transport in Hilbert space?
A naive way of forcing the analogy between spacetime parallel transport and
the path integral or Dirac phase factor is to place an $i$ in the exponent of
the holonomy map
(\ref{Ucirc}), which makes the connections imaginary.  This is equivalent to
performing an analytic continuation into imaginary time.  One then finds that for
the circular orbits
the exponential and oscillatory solutions are reversed with
oscillatory solutions in the region $r < r_{crit}$ and exponential
solutions for $ r > r_{crit}$.  The holonomy map (\ref{Ucirc})
for circular
orbits  remains the same,
with hyperbolic functions replacing trignometric ones
and $S$ replaced by $\omega$.
As noted by \cite{Bollini}, the same result is obtained by considering
the region $r < 2M$, where the roles of space and time are reversed.  We have
not investigated the consequences of classical imaginary-time holonomy in
detail, but it does not appear to mimic known quantum effects.\\

The next natural thought for combining the spacetime and quantum descriptions is
to consider a Kaluza-Klein model, in which the gauge and gravitational
potentials are placed
on an equal footing.  However, when one does
 this and compactifies to four dimensions, one
finds solutions similar to the Reissner-Nordstr\"om black hole.  That is, one
merely gets another connection coefficient from the fifth dimension, which
adds another term to the holonomy frequency $\omega$, just
as we found in the previous section.  Indeed, this does
couple the electromagnetic potential to the holonomy through the charge, but
not in a qualitatively new way; it does not in any sense change a spacetime
vector into a Hilbert space vector.  Apparently a more sophisticated
prescription is necessary to get qualitatively new behavior.
Those advanced so far by the above authors generally involve constructing
Hilbert bundles over spacetime; however at present no particular
formulation
appears to be universally accepted. Exploration of quantum parallel transport
thus remains a promising territory for research.

\section{Acknowlegements}

We would like to thank the referees for helpful comments. This work is partially
funded by the NRF (South Africa). J.M. acknowledges support from a Sainsbury
fellowship.

\end{document}